\documentclass[doublecol]{epl2}
\usepackage{amsmath}
\usepackage{graphicx}

\usepackage{color}

\title{Evaluating linear response in active systems 
with no perturbing field}
\shorttitle{Linear response in active systems with no perturbing field}

\author{Grzegorz Szamel\inst{1,2}}
\shortauthor{Grzegorz Szamel}
\institute{
  \inst{1} Department of Chemistry, 
Colorado State University, Fort Collins, CO 80523, USA \\
  \inst{2} Laboratoire Charles Coulomb, UMR 5221 CNRS,
Universit{\'e} de Montpellier, Montpellier, France
}

\date{\today}


\abstract{
We present a method for the evaluation of time-dependent linear response functions
for systems of active particles propelled by a persistent
(colored) noise from unperturbed simulations.  
The method is inspired by the Malliavin weights sampling method proposed by 
Warren and Allen [Phys. Rev. Lett. \textbf{109}, 250601 (2012)] for 
out-of-equilibrium systems of
passive Brownian particles. We illustrate our method by evaluating two linear response
functions for a single active particle in an external harmonic potential.
As an application, we calculate the time-dependent mobility
function and an effective temperature, defined through the Einstein 
relation between the self-diffusion and mobility coefficients, 
for a system of many active particles 
interacting via a screened-Coulomb potential. We find that this effective temperature 
decreases with increasing persistence time of the self-propulsion. Initially, for not 
too large persistence times, it changes rather slowly, but then it decreases markedly 
when the persistence length of the self-propelled motion becomes comparable with the 
particle size. 
}

\begin{document}

\maketitle 

\section{Introduction}
One is often interested in the time-dependent response of a many-particle system to 
an external perturbation. In particular, long-time limits of functions
describing responses to weak, time-independent perturbations give 
linear susceptibilities and linear transport coefficients. For systems in
thermal equilibrium, calculating these 
time-dependent linear response functions is relatively 
easy since they are related, via fluctuation-dissipation relations, 
to time-dependent correlation functions evolving with unperturbed
dynamics \cite{Chandler,HansenMcDonald}.  

Fluctuation-dissipation relations are, in general, not valid for non-equilibrium systems.
In particular, they are not valid for systems under an external drive 
(\textit{e.g.} sheared systems) or systems driven internally (\textit{e.g.} 
systems consisting of self-propelled/active objects). 
Thus, in principle, in order to calculate linear response in non-equilibrium systems
one has to run simulations at a finite value of the perturbation and approximate
the susceptibility by a finite difference (with due care given to 
the perturbation being weak enough so that one is in the linear response regime).
Notably, due to the lack of time-translational invariance, evaluating
time-dependent response functions requires performing many independent simulations in
order to obtain statistically significant results. This makes such calculations
computationally very expensive.

Fortunately, at least for some cases, methods have been developed which allow one to 
calculate linear response functions from simulations of un-perturbed non-equilibrium
systems. Chatelain \cite{Chatelain} and Ricci-Tersenghi \cite{Ricci} introduced closely 
related methods to calculate linear response functions in aging Ising spin systems 
evolving with Monte Carlo dynamics. These methods were re-derived
within a more general approach and compared with other no-field methods by
Corberi \textit{et al.} \cite{Corberi}. Berthier \cite{Berthier2007} derived a similar
method for aging glass-forming fluids evolving with Monte Carlo dynamics.
Finally, Warren and Allen \cite{WarrenAllen} presented a general approach for calculating
linear response functions in systems of interacting particles evolving with 
continuous-time Brownian dynamics. Warren and Allen placed their method (and 
earlier approaches of Refs. \cite{Chatelain,Ricci,Berthier2007}) in the context of
Malliavin weight sampling used in quantitative finance to evaluate price 
sensitivities of derivative securities (``Greeks'') \cite{ChenGlasserman}. 

In this Letter we present a method for calculating time-dependent
linear response functions for a class of non-equilibrium \emph{active} systems
from simulations without any perturbing field. The particles comprising these
systems move under the influence of self-propulsion. We model the 
self-propulsion as a persistent (colored) noise. The presence
of a finite persistence time of the noise requires a non-trivial
generalization of the Malliavin weight sampling method derived by Warren and Allen.

One important motivation for our method is to model active microrheology experiments
\cite{MizunoWilhelmAhmed} which are used to describe mechanical properties of
active biological systems. These experiments monitor frequency-dependent linear
response functions which can be obtained from the time-dependent ones via Fourier
transforms. More generally, our method opens the way to efficient calculations of
sensitivities to changes of external parameters/conditions
of stochastic processes evolving under the influence of colored noises.

We apply our method to a model system consisting of active particles 
with the self-propulsion evolving according to the
Ornstein-Uhlenbeck process. Originally, we introduced this system \cite{Szamel2014} 
as a continuous-time version of a Monte Carlo model proposed by 
Berthier \cite{Berthier2014}. The same model was independently introduced by 
Maggi \textit{et al.} \cite{Maggi2015}. It has recently been studied by 
Fodor \textit{et al.} \cite{Fodor2016} and termed the active Ornstein-Uhlenbeck particles
(AOUPs) model. We note that an approximate mapping has been
proposed \cite{Farage2015} between the AOUP system and the standard
active Brownian particles model \cite{tenHagen}.

A single component AOUP system is characterized by three parameters, the 
number density, single-particle effective temperature (which determines the
long-time diffusion coefficient of an isolated particle) and persistence time
of the self-propulsion. In the limit of vanishing persistence time an AOUP system
becomes equivalent to a thermal Brownian system at the temperature equal to the
single-particle effective temperature. Interestingly, 
Fodor \textit{et al.} found that for a range
of persistence times an AOUP system can be approximated by an equilibrium
system with an effective, persistence time-dependent 
potential \cite{Maggi2015,Farage2015}. 

In the following, we  
state the main result, illustrate it using two analytically solvable examples, present a
non-trivial application, and close with some discussion.
The derivation of the main result is outlined in the Appendix.

\section{Main result} 
To simplify the notation, we will discuss a single self-propelled particle
evolving under the influence of an external force. The generalization to a 
system of many interacting self-propelled particles is straightforward.
We will write the equations of motion in the form used in Ref. 
\cite{Flenner2016}, which is more consistent with equations of motion
used by Fodor \textit{et al.} \cite{Fodor2016} than the original equations
of motion of Ref. \cite{Szamel2014},
\begin{eqnarray}
\label{dxdt}
\dot{x} &=& \xi_0^{-1}\left( F(x) + f \right) \\
\label{dfdt}
\tau_p \dot{f} &=& - f + \eta.
\end{eqnarray}
Here $F(x)$ is the external force, $f$ is the self-propulsion, $\xi_0$ is the friction
coefficient of an isolated particle, $\tau_p$ is the persistence time of the
self-propulsion, and $\eta$  a Gaussian white noise with zero mean and
variance $\left<\eta(t) \eta(t')\right>_{\text{noise}} =
2 \xi_0 T_{\mathrm{eff}}^{\mathrm{sp}} \delta(t-t')$, where
$\left< ... \right>_{\text{noise}}$
denotes averaging over the noise distribution, and $T_{\mathrm{eff}}^{\mathrm{sp}}$ is 
the single-particle effective temperature. Without the external force, the long-time
motion of the particle evolving according to Eqs. (\ref{dxdt}-\ref{dfdt}) is
diffusive with the diffusion coefficient $D_0=  T_{\mathrm{eff}}^{\mathrm{sp}}/\xi_0$ 
(we use the system of units such that the Boltzmann constant is equal to 1).

The problem that we want to address 
can be formulated as follows. Let's assume that at $t=0$ the system is 
in the stationary state and then the external force changes, 
$F(x) \rightarrow F_\lambda(x)$.
We would like to evaluate the linear response of a function of the particle's 
position $\Phi(x)$, to this change. In other words, we are interested in
$\frac{d}{d\lambda}\left<\Phi(x(t))\right>_{\lambda}$.
Here $\left< ... \right>_{\lambda}$ denotes averaging for the system prepared at $t=0$ 
in the steady state corresponding to force $F(x)$, and then evolving for $t>0$ under the 
influence of modified force $F_\lambda(x)$. 
All the derivatives with respect to 
$\lambda$ are calculated at $\lambda=0$, corresponding
to the un-perturbed evolution. In the following, $\left< ... \right>$ denotes the 
un-perturbed, steady state average.  

The main result of this Letter is that 
$\frac{d}{d\lambda}\left<\Phi(x(t))\right>_{\lambda}$ can be evaluated
as a weighted average over unperturbed dynamics,
\begin{eqnarray}\label{main}
\frac{d}{d\lambda}\left<\Phi(x(t))\right>_{\lambda} &=& 
\left<\Phi(x(t))\left(q(t)+p(t)\right)\right>
\nonumber \\ &&  
+ \tau_p \left<\dot{\Phi}(x(t))q(t)\right>.
\end{eqnarray}
In Eq. (\ref{main}), $q(t)$ and $p(t)$ are Malliavin-like weighting
functions that evolve according to the following equations of motion,
\begin{eqnarray}\label{dqdt}
\dot{q} &=& \frac{1}{2\xi_0^2 T_{\mathrm{eff}}^{\mathrm{sp}}}
\frac{\partial F_\lambda(x)}{\partial\lambda} \eta
\\
\label{dpdt}
\dot{p} &=& \frac{1}{2\xi_0^3 T_{\mathrm{eff}}^{\mathrm{sp}}} 
\tau_p \left(F(x)+ f \right)
\frac{\partial^2 F_\lambda(x)}{\partial x\partial\lambda} \eta
\end{eqnarray}
with initial conditions $q(0)=0=p(0)$. 

We note that by taking the $\tau_p\to 0$ limit while keeping 
$T_{\mathrm{eff}}^{\mathrm{sp}}$ constant, the equations of motion 
(\ref{dxdt}-\ref{dfdt}) become equivalent to the Langevin equation describing
a Brownian particle moving under the influence of an external force with thermal
noise determined by $T=T_{\mathrm{eff}}^{\mathrm{sp}}$. Correspondingly, in the same
limit both the weight $p(t)$ and the second average at the right-hand-side of
Eq. (\ref{main}) vanish, and our main result, Eq. (\ref{main}), 
becomes equivalent to the main result of Warren and Allen, 
Eq. (3) of Ref. \cite{WarrenAllen}. 

\section{Examples} We consider a self-propelled particle in a harmonic
potential.
The equations of motion read
\begin{eqnarray}
\label{dxdth0}
\dot{x} &=& \xi_0^{-1} \left( -k x + f \right), \\
\label{dvdth0}
\tau_p \dot{f} &=& - f + \eta.
\end{eqnarray}

In the first example, we perturb the system by a constant force $\lambda_1$. This 
amounts to the substitution $-kx \rightarrow -kx + \lambda_1$ at the 
right-hand-side of Eq. (\ref{dxdth0}).  
The constant force shifts the average position of the particle away from the 
center of the harmonic potential, and the most interesting linear response
is the change of the average position of the self-propelled particle at
time $t$ after the perturbation was turned on, 
$d \left<x(t)\right>_{\lambda_1}/d\lambda_1$.

\begin{figure}
\includegraphics[width=3in]{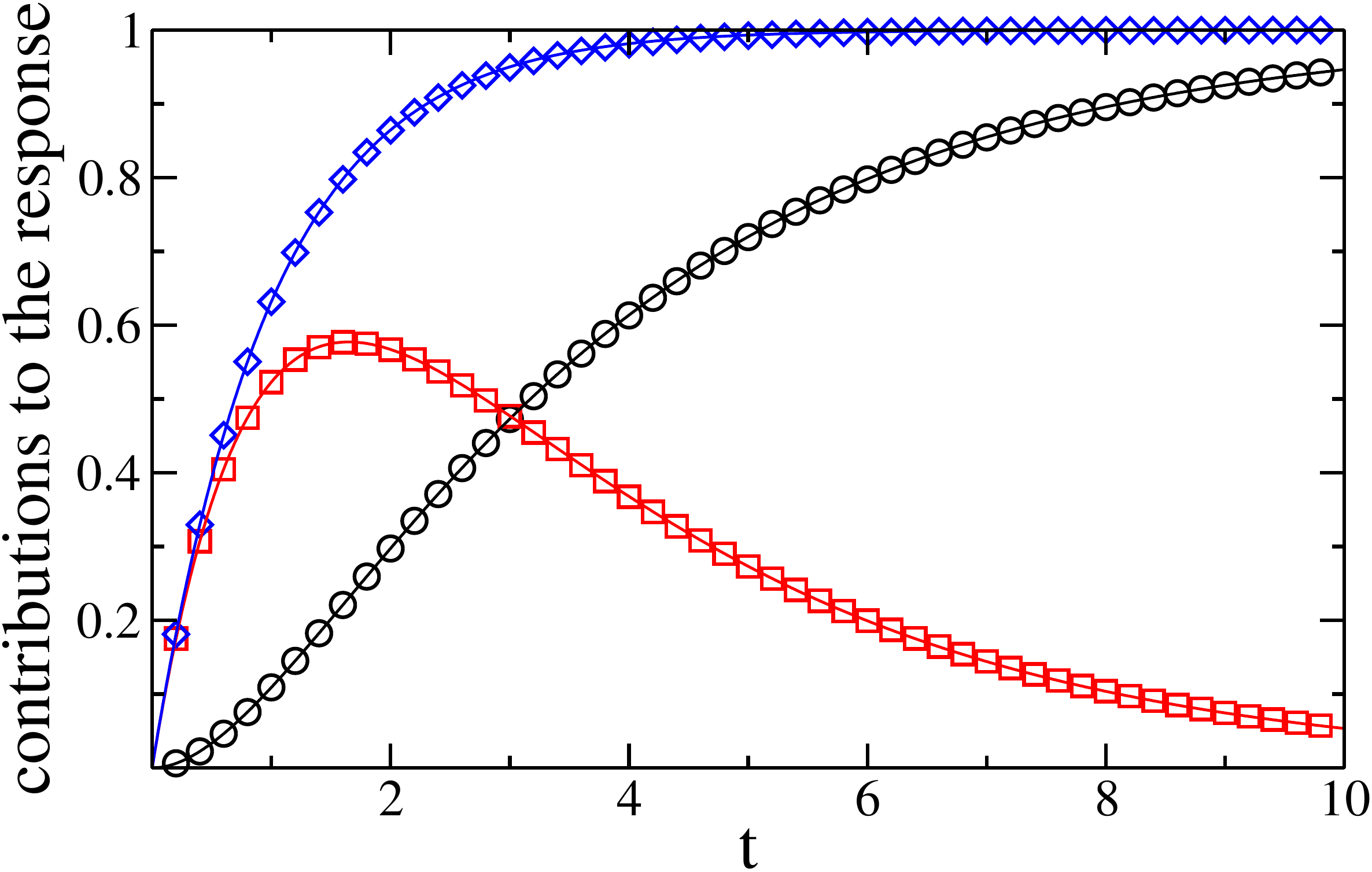}
\caption{\label{harmconstforce} 
The  response of the average position to the constant force 
perturbation, $d \left<x(t)\right>_{\lambda_1}/d\lambda_1$.
Symbols: results of numerical simulations of Eqs. (\ref{dxdth0}-\ref{dqdth}) 
for $k=1$ and $\tau_p=3$.
Circles: $\left<x(t)q_1(t)\right>$, squares: 
$\tau_p\left<\dot{x}(t)q_1(t)\right>$, diamonds: complete response. 
The lines show the analytical formulas, 
Eqs. (\ref{response1a}-\ref{response1}).}
\end{figure}

With the constant force perturbation
$\partial F_{\lambda_1}/\partial\lambda_1 = 1$ and 
$\partial^2 F_{\lambda_1}/\partial x\partial\lambda_1 = 0$, and thus
the equation of motion for weight $q_1(t)$ reads 
\begin{eqnarray} \label{dqdth}
\dot{q}_1 =  \left(2\xi_0^2 T_{\mathrm{eff}}^{\mathrm{sp}} \right)^{-1}\eta
\end{eqnarray}
and weight $p_1(t)$ vanishes.
Eqs. (\ref{dxdth0}-\ref{dqdth}) can be integrated,
\begin{eqnarray}\label{xh}
&& x(t) = x(0) e^{-k t/\xi_0} 
\\ \nonumber && +\frac{f(0)}{k/\xi_0-1/\tau_p} 
\left(e^{-t/\tau_p} - e^{-kt/\xi_0}\right)
\\ \nonumber && 
+ \tau_p^{-1} \int_0^t dt' e^{-k (t-t')/\xi_0} 
\int_0^{t'} dt'' e^{-(t'-t'')/\tau_p}\eta(t''),
\\ \label{qh}
&& q_1(t) = \left(2\xi_0^2 T_{\mathrm{eff}}^{\mathrm{sp}} \right)^{-1} 
\int_0^t dt' \eta(t').
\end{eqnarray}
To calculate 
$d \left<x(t)\right>_{\lambda_1}/d\lambda_1$, we need to evaluate weighted
averages $\left<x(t)q_1(t)\right>$ and $\tau_p \left<\dot{x}(t)q_1(t)\right>$. 
Using Eqs. (\ref{xh}-\ref{qh}) we obtain,
\begin{eqnarray}\label{response1a}
\left<x(t)q_1(t)\right> &=& 
\frac{1- e^{-k t/\xi_0}}{k} - \frac{e^{-t/\tau_p} - e^{-k t/\xi_0}}{k - \xi_0/\tau_p}
\\
\label{response1b}
\tau_p \left<\dot{x}(t)q_1(t)\right> &=& 
\frac{e^{-t/\tau_p} - e^{-k t/\xi_0}}{k - \xi_0/\tau_p}
\end{eqnarray}
Thus,
\begin{eqnarray}\label{response1}
\frac{d \left<x(t)\right>_{\lambda_1}}{d\lambda_1} = 
\left<x(t)q_1(t)\right> + \tau_p \left<\dot{x}(t)q_1(t)\right> =
\frac{1- e^{-k t/\xi_0}}{k},
\end{eqnarray}
which agrees with the result obtained by solving and then averaging the perturbed
equations of motion.
In Fig. \ref{harmconstforce} we compare results of 
numerical simulations of Eqs. (\ref{dxdth0}-\ref{dqdth}) with analytical
formulas (\ref{response1a}-\ref{response1}). On the scale of the
figure, the simulation results are indistinguishable from the analytical predictions.

We note that although for the constant force perturbation 
weighting function $p_1(t)$ vanishes, to get the linear response we 
need to include the term $\tau_p \left<\dot{x}(t)q_1(t)\right>$, which 
implies that even in this case our method is different from that of Warren and
Allen \cite{WarrenAllen}.

In the second example, we perturb the force constant in Eq. (\ref{dxdth0}),
$k\rightarrow k-\lambda_2$. The change of the force constant does not
change the average position of the self-propelled particle. Instead, it changes the 
spatial extend of the particle's steady-state distribution, and the most 
interesting linear response function measures the change of the average 
square position of the particle at
time $t$ after the perturbation was turned on,
$d \left<x^2(t)\right>_{\lambda_2}/d\lambda_2$.
In this case, $\partial F_{\lambda_2}/\partial\lambda_2 = x$ 
and $\partial^2 F_{\lambda_2}/\partial x\partial\lambda_2 = 1$, and thus
equations of motion for weights $q_2(t)$ and $p_2(t)$ read
\begin{eqnarray} \label{dqdth1}
\dot{q}_2 &=&  \left(2\xi_0^2 T_{\mathrm{eff}}^{\mathrm{sp}} \right)^{-1} x \eta
\\
\label{dpdth1}
\dot{p}_2 &=&  \left(2\xi_0^3 T_{\mathrm{eff}}^{\mathrm{sp}} \right)^{-1} 
\tau_p \left(-kx+ f\right) \eta,
\end{eqnarray}
where It\^{o} prescription is implied. To calculate 
$d \left<x^2(t)\right>_{\lambda_2}/d\lambda_2$, we need to evaluate weighted
averages $\left<x^2(t)q_2(t)\right>$, $\left<x^2(t)p_2(t)\right>$, 
and $\tau_p \left<\dot{x^2}(t)q_2(t)\right>
\equiv 2 \tau_p \left<\dot{x}(t)x(t)q_2(t)\right>$. Analytically, it
is convenient to evaluate these averages using the exact equation of motion for
the joint probability distribution of $x$, $f$, $q_2$ and $p_2$ that can be obtained from
Eqs. (\ref{dxdth0}-\ref{dvdth0}) and (\ref{dqdth1}-\ref{dpdth1}). 
The resulting
expressions, consisting of exponential functions, are rather long. Here
we present only the final formula for the linear response,
\begin{eqnarray}\label{response2}
\lefteqn{\!\!\!\!\!\!\!\!\!\!\!\!\!\!\!\! 
\frac{d \left<x^2(t)\right>_{\lambda_2}}{d\lambda_2} =  
\frac{1-e^{-2 k t/\xi_0}}{k} \left<x^2\right> + 
\left[\frac{1-e^{-(k + \xi_0/\tau_p )t/\xi_0}}{k(k+\xi_0/\tau_p)} 
\right. } \nonumber \\ && \left. -\frac{e^{-(k+\xi_0/\tau_p) t/\xi_0} 
-e^{-2k t/\xi_0}}{k(k -\xi_0/\tau_p)}
\right]
\left<xf\right>.
\end{eqnarray}
In Fig. 2 we compare the results of numerical simulations of Eqs. 
(\ref{dxdth0}-\ref{dvdth0}) and (\ref{dqdth1}-\ref{dpdth1}) with the
analytical predictions. Again, the simulation 
and analytical results are indistinguishable.

\begin{figure}
\includegraphics[width=3in]{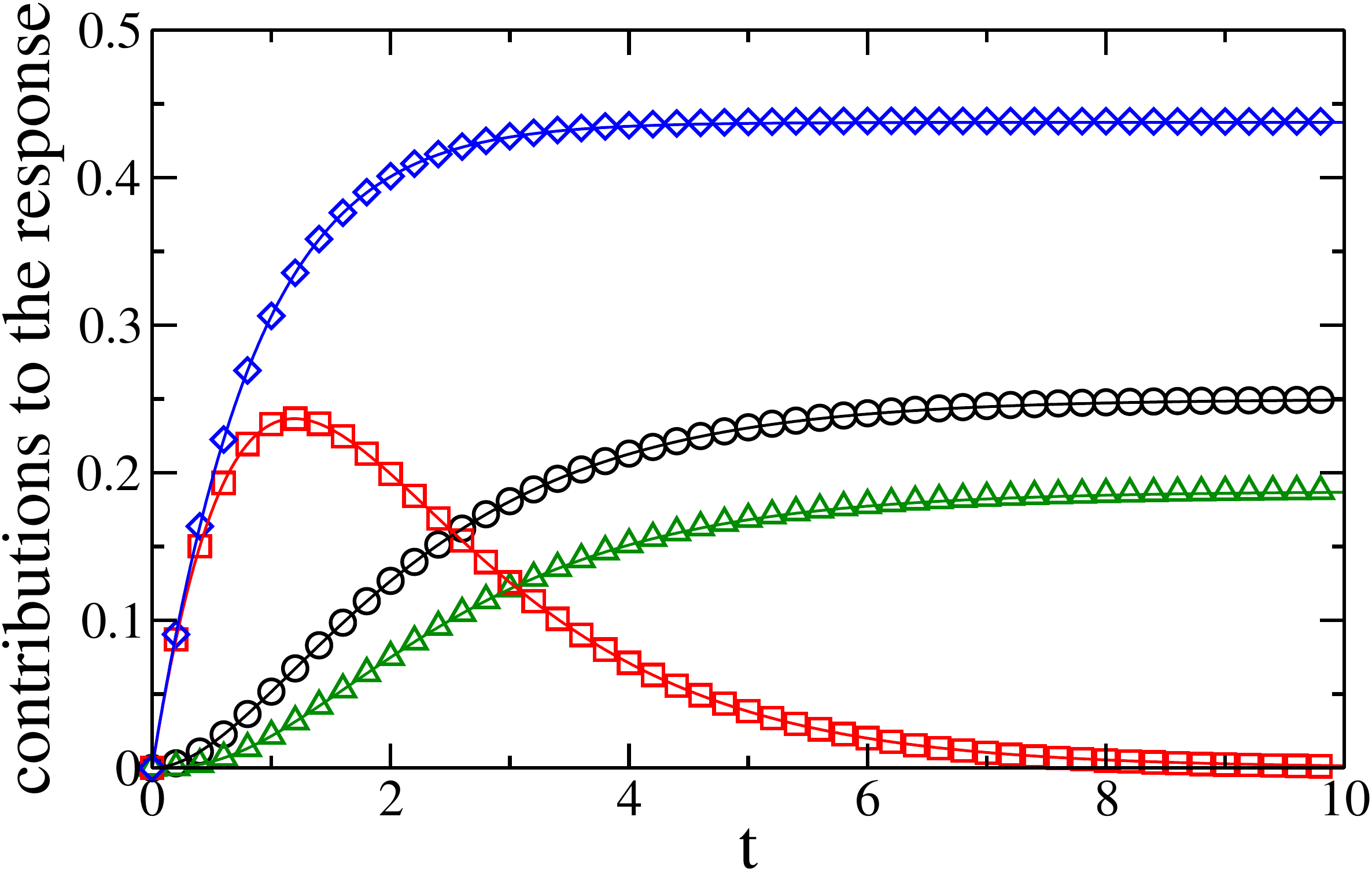}
\caption{\label{harmforceconst} 
The  response of the average square position to a change in the force constant, 
$d \left<x^2(t)\right>_{\lambda_2}/d\lambda_2$.
Symbols: results of numerical simulations of Eqs. (\ref{dxdth0}-\ref{dvdth0}) and
(\ref{dqdth1}-\ref{dpdth1}) for $k=1$ and $\tau_p=3$.
Circles: $\left<x^2(t)q_2(t)\right>$, triangles: $\left<x^2(t)p_2(t)\right>$,
squares: $\tau_p\left<\dot{x^2}(t)q_2(t)\right>$, diamonds: complete response. 
The lines show the analytical predictions.}
\end{figure}

Interestingly, the response of the average position to the constant force 
perturbation, Eq. (\ref{response1}), is independent of the persistence
time of the self-propulsion and is exactly the same as that of a thermal Brownian 
particle in a harmonic potential. In contrast, the response of the average square 
position to a change in the force constant depends on the persistence time. 
In particular, the expressions for the long-time asymptotic responses read
\begin{eqnarray}\label{response1lt}
\lim_{t\to\infty} \frac{d \left<x(t)\right>_{\lambda_1}}{d\lambda_1} &=& \frac{1}{k},
\\ 
\label{response2lt}
\lim_{t\to\infty} \frac{d \left<x^2(t)\right>_{\lambda_2}}{d\lambda_2} &=& 
\frac{ T_{\mathrm{eff}}^{\mathrm{sp}}\left(1+2k\tau_p/\xi_0\right)}
{k^2\left(1+k\tau_p/\xi_0\right)^2}.
\end{eqnarray}
For a thermal Brownian particle in a harmonic potential the long-time
asymptotic responses can be expressed in terms of equilibrium 
correlations, 
$\lim_{t\to\infty} d \left<x(t)\right>_{\lambda_1}^{\mathrm{eq}}/d\lambda_1 = 
\left<x^2\right>^{\mathrm{eq}}/T$ and 
$\lim_{t\to\infty}d\left<x^2(t)\right>_{\lambda_2}^{\mathrm{eq}}/d\lambda_2 = 
\left<x^2\left(x^2 - \left<x^2\right>^{\mathrm{eq}}\right)\right>^{\mathrm{eq}}/(2T)$,
where $\left<\dots\right>^{\mathrm{eq}}_{\lambda_i}$ denotes a perturbed equilibrium 
canonical average at temperature T and 
$\left<\dots\right>^{\mathrm{eq}}$ denotes the un-perturbed equilibrium
canonical average. 
For the self-propelled particle in a harmonic
potential these correlations can be calculated analytically,
\begin{eqnarray}\label{fluct1}
\left<x^2\right> &=& \frac{ T_{\mathrm{eff}}^{\mathrm{sp}}}
{k\left(1+k\tau_p/\xi_0\right)},
\\
\label{fluct2}
\left<x^2\left(x^2 - \left<x^2\right>\right)\right> &=& 
\frac{2\left(T_{\mathrm{eff}}^{\mathrm{sp}}\right)^2}
{k^2\left(1+k\tau_p/\xi_0\right)^2}.
\end{eqnarray}
We note that the right-hand-side of Eq. (\ref{fluct1}) divided by 
$T_{\mathrm{eff}}^{\mathrm{sp}}$ underestimates the response (\ref{response1lt})
and likewise the right-hand-side of Eq. (\ref{fluct2}) divided by 
$2T_{\mathrm{eff}}^{\mathrm{sp}}$ underestimates the response (\ref{response2lt}).

\section{Application} As an application to a system for which exact analytical
formulas cannot be obtained, we evaluated the time-dependent 
mobility of a single particle in a system of interacting self-propelled particles. 
The mobility function describes the change of the position of 
a single particle after the application of a weak force. 
Specifically, at $t=0$ a constant force is applied to particle number 1, 
$F_1(\mathbf{r}_1)\rightarrow F_1(\mathbf{r}_1)+\lambda \mathbf{e}_{\alpha}$. Here
$\mathbf{r}_i$ is the position of particle $i$, $F_i(\mathbf{r}_i)$ is the interparticle
force acting on particle $i$,
$F_i(\mathbf{r}_i) = -\partial_{\mathbf{r}_i} \sum_{l>m} V(r_{lm})$, 
with $V(r)$ being the potential, 
and $\mathbf{e}_\alpha$ is a unit vector in the 
Cartesian direction $\alpha$, $\alpha=x,y,z$. Under the 
influence of the constant force, particle 1 starts moving and the $\alpha$ 
component of its position starts changing in a systematic way,
\begin{eqnarray}
\left<\alpha_1(t)\right>_\lambda = \chi(t) \lambda + o(\lambda),
\end{eqnarray}
where $\chi(t)$ is the mobility function. In the long-time limit,
particle $1$ achieves a constant velocity and then 
$\alpha$ component of its position changes linearly with time.
This allows us to define mobility coefficient $\mu$, 
\begin{eqnarray}\label{smobility}
\left<\alpha_1(t)\right>_\lambda \sim \mu t \lambda \text{ for } t \gg 1.
\end{eqnarray}
In general, $\mu$ depends on the density, the single-particle 
effective temperature and the persistence time. 
Its inverse is the single-particle friction coefficient, $\xi=1/\mu$.

We evaluated the mobility function for a $d=3$ dimensional system 
of $N=1372$ AOUPs interacting via a screened-Coulomb potential,
$V(r) = A \exp\left(-\kappa(r-\sigma)\right)/r$,
with $A=475 T_{\mathrm{eff}}^{\mathrm{sp}} \sigma$
and $\kappa\sigma=24$, at number density $N\sigma^3/V = 0.51$. 
The parameters were chosen in such a way that in the 
limit of vanishing persistence time the present system becomes equivalent to 
a colloidal system that we investigated in the past \cite{Szamel2001}. 
In the following we use reduced units, with $\sigma$ being the unit of length 
and $\sigma^2\xi_0/T_{\mathrm{eff}}^{\mathrm{sp}}$ being the unit of time. 
We note that the system we considered is rather dense and its steady-state
structure factor, for all persistence times investigated, for small wavevectors is 
approximately constant and small. This suggests that this system does not undergo 
a phase separation into a dilute and dense components, at least for the persistence
times investigated.

The most direct application of the approach presented here would be to run an 
un-perturbed simulation and, starting at $t=0$, monitor the weighting function 
$q_{1\alpha}(t)$, which evolves according to equation of motion
\begin{eqnarray}
\dot{q}_{1\alpha} = 
\left(2\xi_0^2 T_{\mathrm{eff}}^{\mathrm{sp}}\right)^{-1} \eta_{1\alpha},
\end{eqnarray}
where $\eta_{i\alpha}$ is the $\alpha$ component of the noise acting on the
self-propulsion of particle $i$. Then, to get the response function
one would need to evaluate 
$\left<\alpha_1(t) q_{ 1\alpha}(t)\right>
+\tau_p\left<\dot{\alpha}_1(t) q_{ 1\alpha}(t)\right>$.
In practice, it is advantageous to monitor $3N$ weighting functions 
corresponding to all particles and all Cartesian directions, and to average
over time origins. This results in the following expression for the 
mobility function,
\begin{eqnarray}
\chi(t) &=& \frac{1}{d N N_{t_0}} \sum_{t_0, \alpha,i}
\left[ \left<\alpha_i(t) (q_{i\alpha}(t+t_0)-q_{i\alpha}(t_0)\right> \right.
\nonumber \\ && \left. + 
\tau_p \left<\dot{\alpha}_i(t) (q_{i\alpha}(t+t_0)-q_{i\alpha}(t_0)\right> \right],
\end{eqnarray}
where $N_{t_0}$ is the number of time origins.

We should emphasize at this point that it is averaging over time origins that
makes it possible to efficiently calculate the response function from a single
unperturbed trajectory.

\begin{figure}
\includegraphics[width=3in]{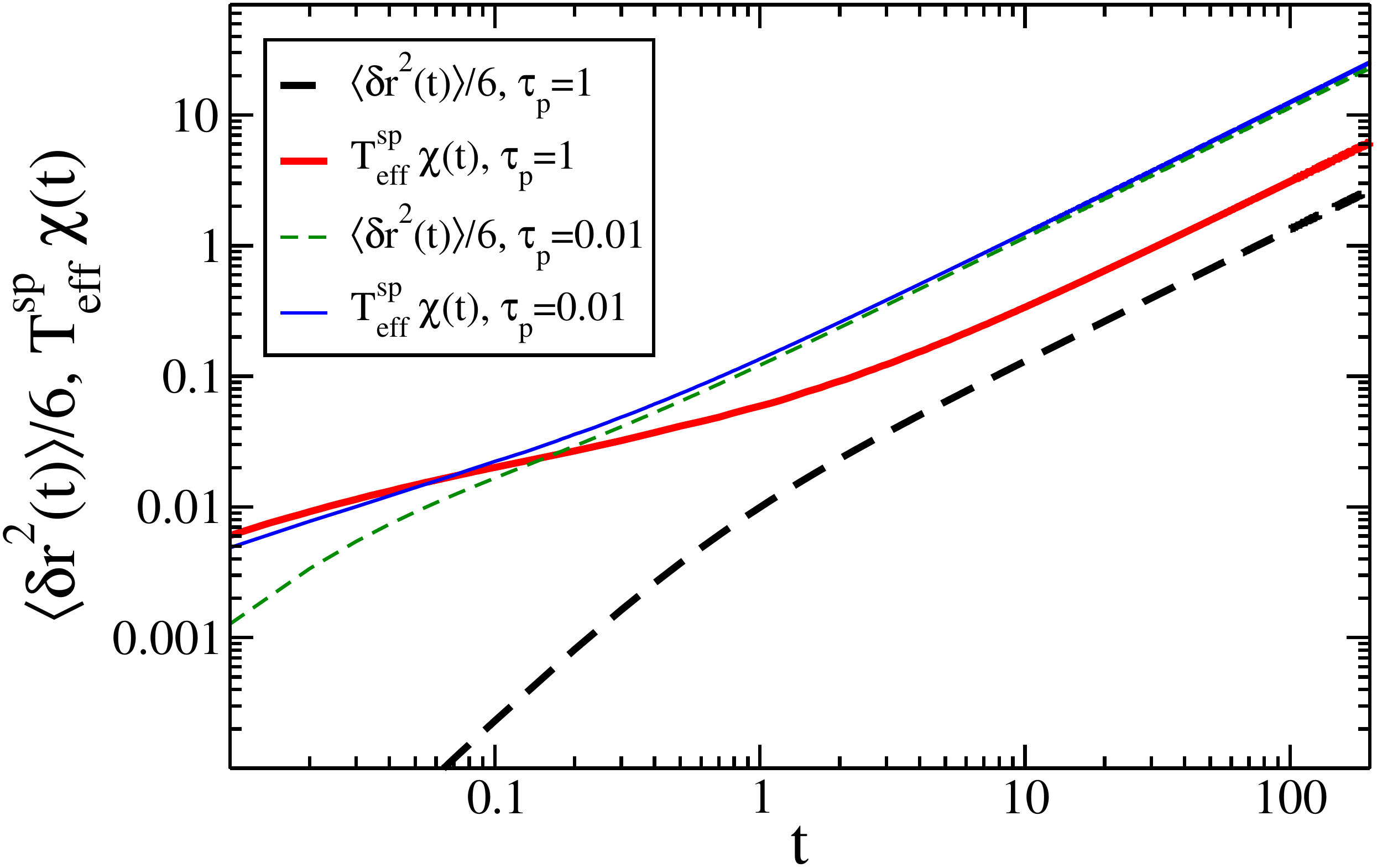}
\caption{\label{yukconstforce} 
Time dependence of $T_{\mathrm{eff}}^{\mathrm{sp}}\chi(t)$ (solid lines)
and MSD/6 (dashed lines). 
Thin lines show $\tau_p=0.01$ and thick lines show $\tau_p=1$.
The figure shows that 
the effective temperature based on the Einstein relation, 
$T_{\mathrm{eff}}^{\mathrm{E}} = D/\mu$, is close 
to $T_{\mathrm{eff}}^{\mathrm{sp}}$
for $\tau_p=0.01$ and is notably smaller than $T_{\mathrm{eff}}^{\mathrm{sp}}$
for $\tau_p=1$.}
\end{figure}

For a system in thermal equilibrium, mobility function $\chi(t)$ is
simply related to the mean-square displacement (MSD),
\begin{eqnarray}
T \chi(t) = \left(2 d \right)^{-1}
\left< \left(\mathbf{r}_i(t)-\mathbf{r}_i(0)\right)^2\right>_{\mathrm{eq}}.
\end{eqnarray}
In the long-time limit the MSD grows as $2d D t$, with $D$ being the self-diffusion
coefficient. Combining definition of the mobility (\ref{smobility}) and 
the asymptotic time-dependence of the MSD we get the Einstein relation,
$T \mu = D$. This relation is, in general, not valid outside of thermal
equilibrium. 

The Einstein relation can, however, be used to
define the following effective temperature 
\begin{eqnarray}\label{TeffE}
T_{\mathrm{eff}}^{\mathrm{E}} = D/\mu.
\end{eqnarray}
Here the superscript $\mathrm{E}$ indicates that $T_{\mathrm{eff}}^{\mathrm{E}}$ 
is defined through the Einstein relation. This effective temperature depends 
on the density, the single-particle effective temperature and the persistence time, and
in general it is different than $T_{\mathrm{eff}}^{\mathrm{sp}}$, except in
the low density or vanishing persistence time limits. We showed previously
that for a sheared Brownian system $T_{\mathrm{eff}}^{\mathrm{E}}$ determines the 
density distribution in a slowly varying external potential beyond linear response
\cite{Szamel2011}. It would be interesting to investigate whether this is also
true for an AOUPs system.

In Fig. \ref{yukconstforce} 
we compare mobility $T_{\mathrm{eff}}^{\mathrm{sp}}\chi(t)$ and the MSD/6
for two values of the persistence
time, $\tau_p=0.01$ and $\tau_p=1$. For short times these two
functions are notably different. In fact, it can be showed that in the short-time
limit $\chi(t)$ grows linearly with time whereas the MSD grows quadratically with
time. In the long time limit both functions grow linearly with time. It can be
seen the long time limits of $T_{\mathrm{eff}}^{\mathrm{sp}}\chi(t)$ and MSD/6 
are very close for $\tau_p=0.01$ 
and markedly different for $\tau_p=1$. This implies that for $\tau_p=0.01$ 
the effective temperature $T_{\mathrm{eff}}^{\mathrm{E}}$
is close to $T_{\mathrm{eff}}^{\mathrm{sp}}$ whereas for $\tau_p=1$ these
two temperatures are different.

\begin{figure}
\centerline{\includegraphics[width=1.5in]{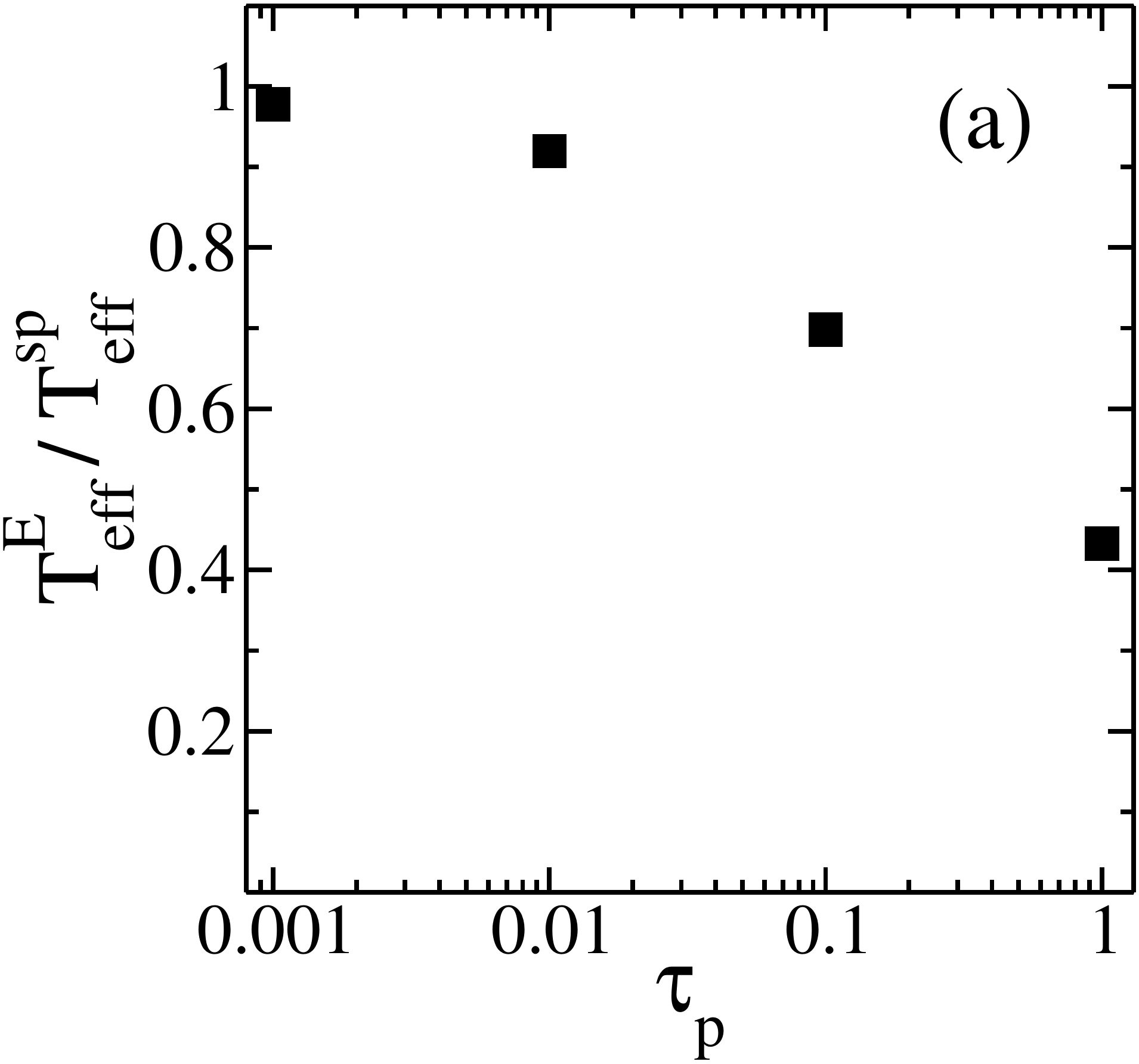}\hspace*{.1in}
\includegraphics[width=1.5in]{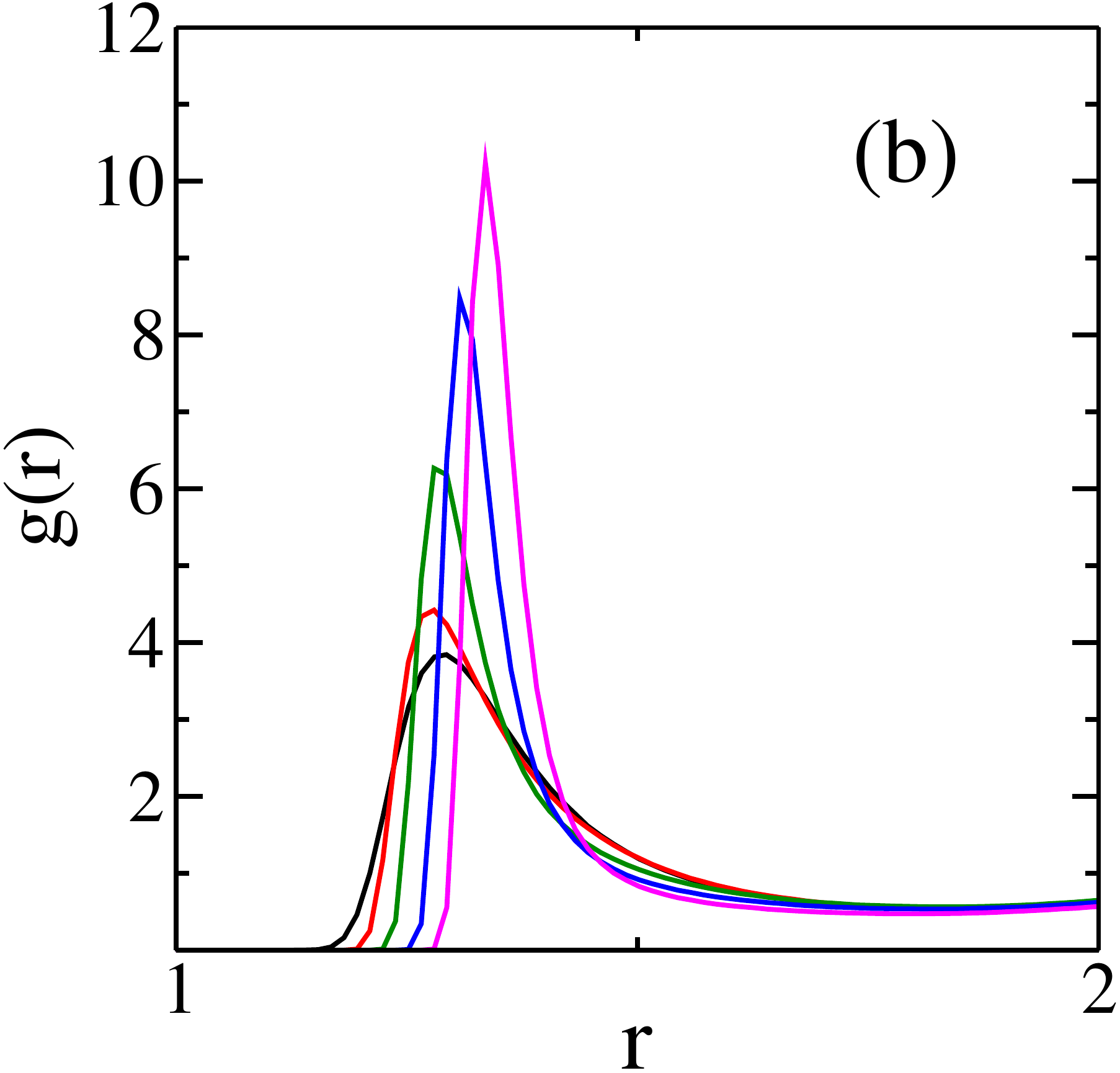}}
\caption{\label{Teffgr} 
(a) Reduced effective temperature defined through the Einstein relation, 
$T_{\mathrm{eff}}^{\mathrm{E}}/T_{\mathrm{eff}}^{\mathrm{sp}}$, 
as a function of persistence time, $\tau_p$.
(b) Dependence of pair distribution function $g(r)$ on
persistence time $\tau_p$. The lines represent, from top to bottom, 
$\tau_p=1,0.1,0.01,0.001$ and the Brownian limit $\tau_p\to 0$.}
\end{figure}

In Fig. \ref{Teffgr}a we show the persistence time dependence of the 
effective temperature defined through the Einstein relation. For a range 
of persistence times $T_{\mathrm{eff}}^{\mathrm{E}}$ changes rather slowly. Then,  
$T_{\mathrm{eff}}^{\mathrm{E}}$ starts changing more rapidly and it reaches 
approximately $T_{\mathrm{eff}}^{\mathrm{sp}}/2$ when the
persistence length of the active motion, 
$l_p\sim\sqrt{T_{\mathrm{eff}}^{\mathrm{sp}}\tau_p/\xi_0}$,
becomes comparable to the range of the potential. 

Interestingly, the dependence of the effective temperature defined through the 
Einstein relation on the persistence time of the self-propulsion is the opposite
of that obtained by Levis and Berthier \cite{LevisBerthier} for the 
system of self-propelled hard disks evolving with Monte Carlo dynamics.
Other evidence of a non-universal dependence of the properties
of active systems on the departure from thermal equilibrium was noted earlier
in Ref. \cite{Flenner2016}. 

In Fig. \ref{Teffgr}b we show the $\tau_p$ dependence of 
the simplest structural quantity, the
pair distribution function $g(r)$ \cite{Chandler,HansenMcDonald}. 
Even for the shortest persistence time investigated, $\tau_p=0.001$, 
the height of the first peak of $g(r)$ is markedly different from its value for
the thermal Brownian system. 

The results showed in Fig. \ref{Teffgr} are consistent with results obtained by 
Fodor \textit{et al.} \cite{Fodor2016}. They found that for 
a range of persistence times the
main effect of the departure from equilibrium is a renormalization of the
interaction potential. This has a profound influence on the local structure
but does not change the effective temperature defined through the 
Einstein relation.

\section{Discussion} We have presented here a method to calculate 
linear response functions for a class of self-propelled systems from
un-perturbed simulations. Our approach generalizes the Malliavin weights 
method to systems evolving under the influence of a persistent noise. 
The method can be easily applied to calculate the response of an AOUPs
system to a periodic in space time-independent potential \cite{LevisBerthier}. 
This will allow us to study the wave-vector dependence of effective temperatures.
We would also like to investigate the response to an externally
imposed shear flow. This will allow us to study the linear viscoelastic
properties of active systems. Finally, it would be interesting to investigate
whether the approach presented here could be generalized to calculate directly
the frequency-dependent response, \textit{i.e.} response to a perturbation
periodic in time \cite{MizunoWilhelmAhmed}, and to calculate directly 
linear response functions for systems of active Brownian particles, without
relying on the mapping procedure proposed in Ref. \cite{Farage2015}.

\section{Acknowledgments}
Most of this work was done when I was visiting Laboratoire Charles Coulomb of
Universit\'{e} de Montpellier. The hospitality of the Laboratoire was greatly 
appreciated. The research in Montpellier is supported
by funding from the European Research Council under the European
Union's Seventh Framework Programme (FP7/2007-2013) /
ERC Grant agreement No 306845. I also gratefully
acknowledge the support of NSF Grant No.~CHE 1213401. I thank E. Flenner for 
comments on the manuscript. 

\section{Appendix: derivation of the main result}  
To evaluate the dependence of $\left<\Phi(x;t)\right>_{\lambda}$ on $\lambda$
we follow the strategy inspired by Sec. 2 of Ref. \cite{ChenGlasserman}. 
Specifically, we write $\left<\Phi(x;t)\right>_{\lambda}$ as
\begin{eqnarray}\label{Phi1}
\lefteqn{
\left<\Phi(x;t)\right>_{\lambda} = }
\nonumber \\ && 
\int dx_N df_N ... dx_0 df_0  \Phi(x_N) 
P_{\lambda}(x_N,f_N|x_{N-1},f_{N-1};\Delta t) 
\nonumber \\ && \times \dots
P_{\lambda}(x_1,f_1|x_0,f_0;\Delta t) P^{ss}(x_0,f_0),
\end{eqnarray}
where for $1<i$ $P_{\lambda}(x_i,v_i|x_{i-1},f_{i-1};\Delta t)$ is the transition
probability over small interval $\Delta t = t/N$, corresponding to the
evolution equations (\ref{dxdt}-\ref{dfdt}) with modified force $F_{\lambda}$,
\begin{eqnarray}\label{transprobmod}
\lefteqn{P_{\lambda}(x_i,f_i|x_{i-1},f_{i-1};\Delta t) = }
\nonumber \\ && 
\delta\left(x_i-x_{i-1}-(F_\lambda(x_{i-1})+f_{i-1})\Delta t/\xi_0\right)
\nonumber \\ && \times 
\frac{
\exp\left( - \frac{(f_i-f_{i-1}+f_{i-1}\Delta t/\tau_p)^2}
{4\xi_0 T_{\mathrm{eff}}^{\mathrm{sp}}
\Delta t/\tau_p^2} \right)
}{\sqrt{4\pi \xi_0 T_{\mathrm{eff}}^{\mathrm{sp}} \Delta t/\tau_p^2}},
\end{eqnarray}
$P_{\lambda}(x_1,v_1|x_0,f_0;\Delta t)$ is the transition
probability with the un-modified force $F$, 
and $P^{ss}(x_0,f_0)$ is the steady state distribution for the un-perturbed system
(\textit{i.e.} the system evolving under the influence of force $F$). 

At this point it is convenient to change the integration variables for $0<i<N$ from
$(x_i,f_i)$ to $(x_i,w_i)$, 
where $w_i = F_\lambda(x_i)+f_i$. Next, one differentiates both
sides of Eq. (\ref{Phi1}) w.r.t. $\lambda$. Then, after some transformations, one
changes the variables back to the original variables 
and one obtains the following equation 
\begin{eqnarray}\label{interm1}
&& \!\!\!\!\!\!\!\!\!\!    \frac{d}{d\lambda}\left<\Phi(x;t)\right>_{\lambda}=
\int dx_N df_N  ... dx_0 df_0  \Phi(x_N) 
\nonumber \\ && \!\!\!\!\!\!\!\!\!\!\!\!  \times
\left[-\frac{(f_N-f_{N-1}+ 
f_{N-1}\Delta t/\tau_p)}{2\xi_0^2 T_{\mathrm{eff}}^{\mathrm{sp}}\Delta t/\tau_p^2}
\left(1-\frac{\Delta t}{\tau_p}\right)
\frac{\partial F_\lambda(x_{N-1})}{\partial\lambda}
\right.\nonumber \\ && \!\!\!\!\!\!\!\!\!\!\!\!  \left. 
+\sum_{i=1}^{N-1} \frac{(f_i-f_{i-1} + 
f_{i-1}\Delta t/\tau_p)}{2\xi_0^2 T_{\mathrm{eff}}^{\mathrm{sp}}\Delta t/\tau_p^2}
\left(\frac{\partial^2 F_\lambda(x_{i-1})}{\partial x_{i-1}\partial\lambda}
\frac{x_i-x_{i-1}}{\Delta t} \right)\Delta t 
\right.\nonumber \\ && \!\!\!\!\!\!\!\!\!\!\!\! \left.
+ \sum_{i=2}^{N-1} \frac{(f_i-f_{i-1} + 
f_{i-1}\Delta t/\tau_p)}{2\xi_0^2 T_{\mathrm{eff}}^{\mathrm{sp}}\Delta t/\tau_p}
\left(
\frac{\partial F_\lambda(x_{i-1})}{\partial\lambda}\right)\Delta t \right.
\nonumber \\ && \!\!\!\!\!\!\!\!\!\!\!\!  \left. 
+  \frac{(f_1-f_0 + 
f_0\Delta t/\tau_p)}{2\xi_0^2 T_{\mathrm{eff}}^{\mathrm{sp}}\Delta t/\tau_p^2}
\frac{\partial F_\lambda(x_0)}{\partial\lambda} \right]
\nonumber \\ && \!\!\!\!\!\!\!\!\!\!\!\!  \times 
P(x_N,f_N|x_{N-1},f_{N-1};\Delta t) \dots 
P(x_1,f_1|x_0,f_0;\Delta t) 
\nonumber \\ && \!\!\!\!\!\!\!\!\!\!\!\!  \times P^{ss}(x_0,f_0).
\end{eqnarray}
Then, one uses an identity which follows from the time-independence of a the steady 
state distribution, 
\begin{eqnarray}\label{interm2}
&& \int dx_N df_N ... dx_0 df_0  
\frac{\Phi(x_N)-\Phi(x_{N-1})}{\Delta t} 
\nonumber \\ && \times \sum_{i=1}^{N-1} \frac{(f_i-f_{i-1} + 
f_{i-1}\Delta t/\tau_p)}{2\xi_0^2 T_{\mathrm{eff}}^{\mathrm{sp}}/\tau_p^2}
\frac{\partial F_\lambda(x_{i-1})}{\partial\lambda}
\nonumber \\ && \times 
P(x_N,f_N|x_{N-1},f_{N-1};\Delta t) ... 
P(x_1,f_1|x_0,f_0;\Delta t) 
\nonumber \\ && \times P^{ss}(x_0,f_0)
=  
\int dx_N df_N  ... dx_0 df_0 \Phi(x_{N})
\nonumber \\ && \left[ - \frac{(f_{N}-f_{N-1} + 
f_{N-1}\Delta t/\tau_p)}{2\xi_0^2 T_{\mathrm{eff}}^{\mathrm{sp}}\Delta t/\tau_p^2}
\frac{\partial F_\lambda(x_{N-1})}{\partial\lambda}
\right.\nonumber \\ && \left. 
+ \frac{(f_1-f_0 + 
f_0\Delta t/\tau_p)}{2\xi_0^2 T_{\mathrm{eff}}^{\mathrm{sp}}\Delta t/\tau_p^2}
\frac{\partial F_\lambda(x_0)}{\partial\lambda}\right]
\nonumber \\ &&
\times P(x_N,f_N|x_{N-1},f_{N-1};\Delta t) \dots
P(x_1,f_1|x_0,f_0;\Delta t)
\nonumber \\ &&
\times P^{ss}(x_{0},f_{0})
\end{eqnarray}
Combining Eqs. (\ref{interm1}) and (\ref{interm2}) 
one arrives at the final Eq. (\ref{final}),
\begin{eqnarray}\label{final}
&&
\frac{d}{d\lambda}\left<\Phi(x;t)\right>_{\lambda}=
\int dx_N df_N ... dx_0 df_0 
\nonumber \\ && \left\{ \Phi(x_N)
\left[ \sum_{i=2}^{N}\frac{(f_i-f_{i-1} + 
f_{i-1}\Delta t/\tau_p)}{2\xi_0^2 T_{\mathrm{eff}}^{\mathrm{sp}}/\tau_p}
\frac{\partial F_\lambda(x_{i-1})}{\partial\lambda} \right. \right.
\nonumber \\ && \left. \left.
+ \sum_{i=1}^{N-1}\frac{(f_i-f_{i-1} + 
f_{i-1}\Delta t/\tau_p)}{2\xi_0^2 T_{\mathrm{eff}}^{\mathrm{sp}}/\tau_p^2}
\frac{\partial^2 F_\lambda(x_i)}{\partial x_i\partial\lambda}
\frac{x_i-x_{i-1}}{\Delta t}\right] 
\right. \nonumber \\ && \left. + 
\frac{\Phi(x_N)-\Phi(x_{N-1})}{\Delta t} \sum_{i=1}^{N-1} \frac{(f_i-f_{i-1} + 
f_{i-1}\Delta t/\tau_p)}{2\xi_0^2 T_{\mathrm{eff}}^{\mathrm{sp}}/\tau_p^2}
\right. \nonumber \\ && \left. \times \frac{\partial F_\lambda(x_{i-1})}{\partial\lambda}
\right\}
\times P(x_N,f_N|x_{N-1},f_{N-1};\Delta t) \dots
\nonumber \\ && \times P(x_1,f_1|x_0,f_0;\Delta t) P^{ss}(x_0,f_0).
\end{eqnarray}
In Eqs. (\ref{interm1}-\ref{final}) 
$P(x_i,f_i|x_{i-1},f_{i-1};\Delta t)$ is the transition probability
corresponding to the un-perturbed evolution. It has the same form as 
the  transition probability (\ref{transprobmod}) but the force is the
unperturbed force $F(x_{i-1})$. 

Assuming that the $\Delta t\to 0$ limit can be taken, we get our main
result, Eq. (\ref{main}), with weights $q(t)$ and $p(t)$ evolving
according to Eqs. (\ref{dqdt}-\ref{dpdt}).

\end{document}